\documentclass[ floatfix,twocolumn]{revtex4}
\usepackage{amsmath}
\usepackage{color}
\usepackage{amssymb}
\usepackage{graphicx}
\usepackage[utf8]{inputenc}
\usepackage[english, russian]{babel}
\usepackage{textcomp}
\usepackage{color}
\usepackage{esint}

\begin{document}

\title{ 
URhGe - Altermagnetic  Ferromagnet}

\author{V.P.Mineev}
\affiliation{Landau Institute for Theoretical Physics, 142432 Chernogolovka, Russia}

\begin{abstract}
It is well known that the anomalous Hall effect in ferromagnetic and strongly paramagnetic metals  in addition to electron skew scattering on impurities is determined by internal mechanism linked to 
the Berry curvature, a quantum-mechanical property of the  electron states of a perfect crystal.
Experimentally, however, it has been established that the Berry curvature does not play any role in the Hall resistance of the ferromagnet URhGe. URhGe is so called altermagnetic ferromagnet 
which  crystal symmetry    includes  operation of time inversion only in combination with rotations and reflections. 
The explanation for strictly zero Berry curvature of electronic states in this material  lies in the non-symmorphic symmetry of its crystal lattice.

\end{abstract}
\date{\today}
\maketitle

Uranium metallic compound URhGe possessing many peculiar properties during long time attracts attention of condensed matter community ( see experimental survey \cite{Aoki2019} and recent theoretical paper \cite{Mineev2024} and references wherein). It has orthorhombic crystal structure and the anisotropic magnetic properties. Below Curie temperature $T_c$ = 9.5K URhGe transfers to ferromagnetic state with magnetic moment along c-crystallographic direction and then at $T_{sc}\approx$ 0.25K becomes superconducting. The Hall effect measurements in URhGe has been reported in the paper by D.Aoki and co-authors \cite{Aoki2014}.

The Hall resistivity $\rho_{xy}$ in ferromagnetic and strongly paramagnetic substances is described by
\begin{equation}
\rho_{xy}=R_0H+R_sM,
\end{equation}
where $R_0$ is the normal Hall coefficient and the second term is attributed to the anomalous Hall effect with the magnetisation $M$.  
The anomalous Hall effect originates from extrinsic part determined by skew scattering, side jump scattering and intrinsic part 
determined by the Berry curvature $\Omega_{xy}$ 
being internal property of band structure of pure crystal \cite{Sinitsyn2005,Nagaosa2010}. 
To extract the Hall coefficient $R_0$  the Hall resistivity data for the field direction along $c$-axis were plotted \cite{Aoki2014} in the form of 
\begin{equation}
\rho_{xy}/H=R_0+c \rho M/H,
\end{equation}
assuming according to the authors \cite{Aoki2014}  that the anomalous Hall effect mainly originates from skew scattering, namely, $R_s \propto \rho $. A good linear dependence of $\rho_{xy}/H$ on $\rho M/H$ is found in the wide temperature range from 300 to 12 K. Similar linear dependence was found in ferromagnetic state in  the temperature interval between 2 and 3 K. 

In fact,  the established linear dependence means that at each temperature anomalous Hall resistivity in URhGe is completely determined by processes of electron scattering including not only skew scatterings but many other processes.  The explanation  of this observation is at the moment absent.
Following to the authors of review \cite{Nagaosa2010} one can only say : “It seems extremely challenging, if not impossible, to develop a predictive theory for these contributions”.

Thus, from results  \cite{Aoki2014} follows
that the intrinsic contribution determined by the Berry curvature plays no role in explanation of  anomalous Hall effect in URhGe. 
Symmetry considerations presented in this Letter allow to prove that the Berry curvature $\Omega_{xy}$ in URhGe is strictly equal to zero. 

URhGe crystallises in the TiNiSi-type orthorhombic crystal structure Pnma where the four U atoms (as well 4 Ge and 4 Rh atoms) are located at four different positions 
\begin{eqnarray}
U_1=(x,\frac{1}{4},z),~~U_2=(\bar x,\frac{3}{4},\bar z),~~~~~~~~~~~~~~\nonumber\\U_3
=(\frac{1}{2}-x,\frac{3}{4},\frac{1}{2}+z),~~U_4=(\frac{1}{2}+x,\frac{1}{4},\frac{1}{2}-z)
\end{eqnarray}
 in elementary cell with lattice parameters $(a,b,c)$ \cite{Tran1998}. Coordinates here are written in units of lattice parameters, $\bar x=-x$ etc, points $x$ and $x+1$ etc indistinguishable. All atoms possess magnetic moment along $c$-direction, either induced by external field in paramagnetic state or spontateous plus induced in ferromagnetic state.
Non-symmorphic symmetry group 
of such crystal consists from 8 elements $g_i$, $i=1,...8$
\begin{eqnarray}
{\bf D}^{ns}_{2h}({\bf C}_{2h})~~~~~~~~~~~~~~~~~~~~~~~~~\nonumber\\
=(E, I,RC_{2y},R\sigma_y,t_dRC_{2x},t_dR\sigma_x,t_{xz}C_{2z},t_{xz}\sigma_z)
\label{G}
\end{eqnarray}
Here, $C_{2x}, C_{2y},C_{2z}$  are rotations on the angle $\pi$ around axis $(x,y,z)$ directed along crystallographic axis $(a,b,c)$, $\sigma_x,\sigma_y,\sigma_z$ etc are the reflections in planes perpendicular to these axis, $R$ is the operation of time reflection and $I$ is the operation of space inversion. Operation 
$t_d=\{(x,y,z)\to(x+\frac{1}{2},y+\frac{1}{2},z+\frac{1}{2})\}$ and $t_{xz}=\{(x,y,z)\to(x+\frac{1}{2},y,z+\frac{1}{2})\}$ correspond to shift on half period along the cell space diagonal and on half period in plane $(xz)$. Under operations of symmetry atomic positions transfer to each other as indicated in the following table (see also \cite{Shick2002}). 
\bigskip
 
 \begin{tabular}{|c|c|c|c|c|}
\hline
 $g_i$& $~U_1~$&$~U_2~$ &$~U_3~$&$~U_4~$\\
 \hline
 ~E~ & $U_1$&$U_2$ &$U_3$&$U_4$\\
\hline
 $I$&  $U_2$&$U_1$ &$U_4$&$U_3$ \\
 \hline
 ~$RC_{2y}$~& ~$U_2$&$U_1$ &$U_4$&$U_3$ \\
  \hline
 ~$R\sigma_y$~& $U_1$&$U_2$ &$U_3$&$U_4$\\
  \hline
  ~$t_dRC_{2x}$~& $U_4$&$U_3$ &$U_2$&$U_1$\\
  \hline
 ~$t_dR\sigma_x$~& $U_3$&$U_4$ &$U_1$&$U_2$\\
   \hline
 ~$t_{xz}C_{2z}$~& $U_3$&$U_4$ &$U_2$&$U_1$\\
   \hline
 ~$t_{xz}\sigma_z$~& $U_4$&$U_3$ &$U_2$&$U_1$\\
   \hline 
 \end{tabular}
 
 \bigskip 
Electron spectrum of a metal such that its group of symmetry $G$ ( magnetic class) contains the operation of time reversal only in combination with rotations or reflections has the form 
\begin{equation}
\varepsilon_{\alpha\beta}({\bf k})=\varepsilon_{\bf k}\delta_{\alpha\beta}+\mbox{\boldmath$\gamma$}_{\bf k}\mbox{\boldmath$\sigma$}_{\alpha\beta},
\label{alt}
\end{equation}
invariant in respect of all the operations of the group $G$.
Here, 
 scalar part of spectrum $\varepsilon_{\bf k}$ and vector part of spectrum $\mbox{\boldmath$\gamma$}_{\bf k}$ both are the even functions of momentum ${\bf k}$.
$\mbox{\boldmath$\sigma$}=(\sigma_x,\sigma_y,\sigma_z)$ are the Pauli matrices in the spin space. 
This type of metals looking like antiferromagnets or ferromagnets in reciprocal space are called {\bf altermagnets} \cite{MineevJL2025,MineevUFN2025}. 
In particular, URhGe is altermagnetic ferromagnet. 

The eigenvalues of the matrix (\ref{alt}) are
\begin{equation}
    \varepsilon_{\lambda}({\bf k})=\varepsilon_{\bf k}+\lambda\gamma,~~~~~~~~ \lambda=\pm,
\label{e3}
\end{equation}
where $\gamma=|\mbox{\boldmath$\gamma$}_{\bf k}|$.
The corresponding eigenfunctions are given by
\begin{eqnarray}
\Psi^+_\alpha({\bf k})=\frac{1}{\sqrt{2\gamma(\gamma+\gamma_z)}}\left (\begin{array} {c}
\gamma+\gamma_z\\
\gamma_+
\end{array}\right),\nonumber\\
~~~~~~~~~~~~\Psi^-_\alpha({\bf k})=\frac{t_+^\star}{\sqrt{2\gamma(\gamma+\gamma_z)}}
\left(\begin{array} {c}
-\gamma_-\\
\gamma+\gamma_z
\end{array}\right),
\label{ps}
\end{eqnarray}
where $\gamma_\pm=\gamma_x\pm i\gamma_y$ and $t_+^\star=-\frac{\gamma_+}{\sqrt{\gamma_+\gamma_-}}$.
The Berry curvature for each band  with $\lambda=\pm$ is \cite{Xiao2010}
\begin{eqnarray}
\Omega^\lambda_{xy}({\bf k})=i\left(\frac{\partial \Psi_\alpha^{\lambda\star}}{\partial k_x}\frac{\partial \Psi_\alpha^{\lambda}}{\partial k_y}
-
\frac{\partial \Psi_\alpha^{\lambda\star}}{\partial k_y}\frac{\partial \Psi_\alpha^{\lambda}}{\partial k_x}\right).
\label{om}
\end{eqnarray}
The interband Berry curvature  is \cite{MineevJL2025}
\begin{equation}
\Omega^{int}_{xy}({\bf k})=\frac{\mbox{\boldmath$\gamma$}}{\gamma^3}
\left(\frac{\partial \mbox{\boldmath$\gamma$}}{\partial k_x}\times
  \frac{\partial \mbox{\boldmath$\gamma$}}{\partial k_y}   \right  ).
\end{equation}

Both intraband and interband contributions to the Berry curvature are determined through the vector part $\mbox{\boldmath$\gamma$}_{\bf k}$ of electron spectrum. For URhGe the vector $\mbox{\boldmath$\gamma$}_{\bf k}$ invariant in respect to all operation of the group (\ref{G}) is
\begin{equation}
\mbox{\boldmath$\gamma$}_{\bf k}=\tilde\gamma_x\sin(k_xa)\sin(k_zc)\hat x -h\hat z,
\label{gamma}
\end{equation}
where $-h\hat z$ is the Zeemann term arising due to spontaneous and also induced magnetisation.
Here, we have  taken into account that  the  operation $t_{d}$  in coordinate space corresponds to
the shift $(\pi/a,\pi/b,\pi/c)$ on half basis vector in the reciprocal space. The operation $t_{xz}$  in coordinate space corresponds
to the shift $(\pi/a,0,\pi/c)$ on half basis vector in the reciprocal space.
The equation (\ref{gamma}) defining the vector  $\mbox{\boldmath$\gamma$}_{\bf k}$ is the simplest possible expression that has the necessary symmetry properties.

The vector $\mbox{\boldmath$\gamma$}_{\bf k}$ in URhGe is independent of $\hat y$ component and also of $k_y$ component of momentum. Hence,
both the contributions to the Berry curvature are strictly equal to zero
\begin{equation}
\Omega^\lambda_{xy}({\bf k})=0,~~~~~~~~\Omega^{int}_{xy}({\bf k})=0.
\end{equation}

On the contrary, in  metal with symmorphic crystal symmetry  
\begin{equation}
{\bf D}_{2h}({\bf C}_{2h})=
(E, C_{2}^{z}, RC_{2}^{x}, RC_{2}^{y},I,\sigma_z,R\sigma_x,R\sigma_y)
\label{vector}
\end{equation}
both contributions to the Berry curvature have finite value.  The vector $\mbox{\boldmath$\gamma$}_{\bf k}$ invariant in respect to all operation of the group (\ref{vector}) is
\begin{eqnarray}
\mbox{\boldmath$\gamma$}_{\bf k}
=\tilde\gamma_x\sin(k_xa)\sin(k_zc)\hat x
+\tilde\gamma_y\sin(k_yb)\sin(k_zc)\hat y -h\hat z,~~
\label{gamma1}
\end{eqnarray}
 the Berry curvature for each band is
\begin{eqnarray}
\Omega^\pm_{xy}({\bf k})=\pm\frac{1}{2}\frac{ab\tilde\gamma_x\tilde\gamma_yc_xc_ys_z^2h}{[(\tilde\gamma_xs_xs_z)^2+(\tilde\gamma_ys_ys_z)^2+h^2]^{3/2}},
\end{eqnarray}
the interband Berry curvature  is 
\begin{equation}
\Omega^{int}_{xy}({\bf k})=-\frac{ab\tilde\gamma_x\tilde\gamma_yc_xc_ys_z^2h}{[(\tilde\gamma_xs_xs_z)^2+(\tilde\gamma_ys_ys_z)^2+h^2]^{3/2}}.
\end{equation}
Here, there was introduced the notations $s_x=\sin(k_xa),s_y=\sin(k_yb),s_z=\sin(k_zc),c_x=\cos(k_xa),c_y=\cos(k_yb)$.

In fact, URhGe has non-symmorphic symmetry ${\bf D}^{ns}_{2h}({\bf C}_{2h})$ and the conclusion about finite Berry curvature  
of electronic states in URhGe
based on symmetry ${\bf D}_{2h}({\bf C}_{2h})$ pointed in \cite{MineevJL2025} is incorrect.

It should be noted, however,   that the theoretical treatment  of superconductivity in URhGe based on symmetry ${\bf D}_{2h}({\bf C}_{2h})$  is nevertheless valid. 
For the superconducting properties is important symmetry of material on spatial scales large in comparison with interatomic distances where 
 a difference between ${\bf D}_{2h}({\bf C}_{2h})$ 
and ${\bf D}^{ns}_{2h}({\bf C}_{2h})$ is  inessential. 
The same is true in respect of treatment of  piezomagnetism in URhGe based on symmetry ${\bf D}_{2h}({\bf C}_{2h})$ made in the paper \cite{Mineev2021} and recently confirmed experimentally by Yonezawa and collaborators \cite{Yonezawa2024}.  

In summary, the paper proves that the strictly zero Berry curvature of electronic states in URhGe is the direct consequence of non-symmorphic crystal symmetry of this material.

 \bigskip

  FUNDING This work was supported by ongoing institutional funding. No additional grants to carry out or direct this particular research were obtained. 
  
\bigskip 
 
 CONFLICT OF INTEREST The author of this work declares that he has no conflicts of interest.

\end{document}